\documentclass[12pt,preprint]{aastex}

\long\def\Ignore#1{\relax}

\newcommand{\Md}        {M_{\rm d}}
\newcommand{\Rd}        {R_{\rm d}}
\newcommand{\Mb}        {M_{\rm b}}

\newcommand\degrees{^\circ}

\newcommand{\Msun}{\mbox{$\rm M_{\odot}$}}
\newcommand{\amp}[1]{\mbox{$A_{\rm #1}$}}
\newcommand{\tme}[1]{\mbox{$\tau_{\rm #1}$}}
\newcommand{\omg}[1]{\mbox{$\Omega_{\rm #1}$}}

\newcommand{\lag}[1]{\mbox{$R_{\rm c,{#1}}$}}

\newcommand{\eg}{e.g.}
\newcommand{\etc}{etc.}
\newcommand{\ie}{i.e.}

\shorttitle{Double-Barred Galaxies From Pseudo-Bulges}
 
\shortauthors{Debattista \& Shen}

\begin{document}
\title{Long-Lived Double-Barred Galaxies From Pseudo-Bulges}

\author{Victor P. Debattista\altaffilmark{1}}
 \affil{Astronomy Department, University of Washington, Box 351580,
Seattle, WA 98195}

\altaffiltext{1}{Brooks Prize Fellow; email: debattis@astro.washington.edu}

\and

\author{Juntai Shen\altaffilmark{2}} 
\affil{McDonald Observatory, The University of
Texas at Austin, 1 University Station, C1402, Austin, TX 78712}
\altaffiltext{2}{Harlan J. Smith Fellow; email: shen@astro.as.utexas.edu}


\begin{abstract}
A large fraction of barred galaxies host secondary bars that are
embedded in their large-scale primary counterparts. These are common
also in gas poor early-type barred galaxies.  The evolution of such
double-barred galaxies is still not well understood, partly because of
a lack of realistic $N$-body models with which to study them.  Here we
report a new mechanism for generating such systems, namely the
presence of rotating pseudo-bulges.  We demonstate with high mass and
force resolution collisionless $N$-body simulations that long-lived
secondary bars can form spontaneously without requiring gas, contrary
to previous claims.  We find that secondary bars rotate faster than
primary ones. The rotation is not, however, rigid: the secondary bars
pulsate, with their amplitude and pattern speed oscillating as they
rotate through the primary bars.  This self-consistent study supports
previous work based on orbital analysis in the potential of two
rigidly rotating bars.  The pulsating nature of secondary bars may
have important implications for understanding the central region of
double-barred galaxies.
\end{abstract}

\keywords{stellar dynamics --- galaxies: bulges --- galaxies:
evolution --- galaxies: formation --- galaxies: kinematics and
dynamics --- galaxies: structure}

\section{Introduction} 
\label{sec:intro}

Double-barred (S2B) galaxies, consisting of a small scale
nuclear/secondary bar (B2) embedded within a large scale primary bar
(B1), have been known for over thirty years \citep[e.g.][]{devauc_75}.
\citet{erw_spa_02} carefully compiled statistics for early-type
optically-barred galaxies from images by both the WIYN telescope and
the {\it Hubble Space Telescope}, and concluded that at least one
quarter of them are double-barred.  The facts that inner bars are also
seen in near-infrared \citep[e.g.][]{mul_etal_97,lai_etal_02}, and
that gas-poor S0s often contain inner bars indicate that most of them
are stellar structures.  S2Bs may play an important role in the
formation and nurture of supermassive black holes (SMBHs).  S2B
galaxies have been hypothesized to be a possible mechanism for driving
gas past the inner Lindblad resonance of B1s, feeding SMBHs and
powering AGN \citep{shl_etal_89}.  S2Bs have also been suggested as a
mechanism for forming SMBHs directly \citep{beg_etal_06}.

Such fueling requires that the B2 and the B1 are dynamically
decoupled.\footnote{In this context, by decoupled we mean only that
$\omg{B2} \neq \omg{B1}$, where \omg{B2} (\omg{B1}) is the pattern
speed of the B2 (B1).}  The random apparent orientations of B1s and
B2s in nearly face-on galaxies points to dynamical decoupling
\citep{but_cro_93, fri_mar_93}.  But images alone cannot reveal how
the two bars rotate through each other.  Kinematic evidence of
decoupling, using either gas or stars, is harder to obtain
\citep{pet_wil_02, sch_etal_2002,moi_etal_04}.  Indirect evidence for
decoupling was suggested by \citet{ems_etal_01} based on rotation
velocity peaks inside the B2s in three S2B galaxies.  Conclusive
direct kinematic evidence for a decoupled B2 was obtained for NGC 2950
by \citet{cor_deb_agu_03} who showed, using the method of
\citet{tre_wei_84}, that its B1 and B2 cannot be rotating at the same
rate.

An important advance in understanding S2B galaxies came from the
development by \citet[][hereafter MS00]{mac_spa_97,mac_spa_00} of the
formalism necessary for studying their orbits.  They introduced the
concept of loops, families of orbits in which particles return to the
same curve, but not the same position, after the two bars return to
the same relative orientation.  MS00 considered two models assuming
that the B2 is more rapidly rotating than the B1: the B2 in
their Model 1 ended near its corotation radius, while in their Model 2
it ended well inside this radius.  MS00 were unable to find loop
orbits supporting the outer parts of the B2 in Model 1 but
succeeded in doing so in the more slowly rotating Model 2.  Using
hydrodynamical simulations of such slowly rotating rigid B2s
\citet{mac_etal_02} found them to be inefficient at driving gas to
small radii.

Such models are not fully self-consistent since, in general, nested
bars cannot rotate rigidly through each other \citep{lou_ger_88}.  In
fact non-solid body rotation was hinted at by the loop orbit
calculations of MS00.
$N$-body simulations provide one route to more self-consistent models
of S2Bs, but until now there existed a paucity of such models.  Most
numerical studies \citep[\eg][]{shl_hel_02, fri_mar_93, eng_shl_04}
required gas to form B2s; for example, \citet{hel_etal_01} formed them
via viscosity-driven instabilities in nuclear gas rings, which lead to
B2s rotating slower than B1s.  But the presence of B2s in a large
fraction of gas-poor early-type galaxies \citep{erw_spa_02,pet_wil_02}
indicates that B2s are not an exclusively gas dynamical phenomenon.
Counter-rotation in stellar disks can lead to decoupled
counter-rotating bars \citep{sel_mer_94, friedl_96, dav_hun_97}, but
such counter-rotation is infrequent \citep{kui_etal_96}.  Only
Rautiainen and collaborators \citep{rau_sal_99, rau_etal_02} have
succeeded in forming long-lived B2s rotating in the same sense as the
B1 in purely collisionless studies.  The mechanism by which the B2s
formed in these however simulations remains unclear.

In light of the increasing evidence that SMBH feedback may play an
important role in galaxy formation \citep{spr_etal_05b} and the
possibility that S2Bs may provoke AGNs, the paucity of self-consistent
$N$-body models of S2Bs is a major hindrance to further theoretical
development.
The time is ripe, therefore, to examine whether unambiguous and
independently rotating nested bars can form in high resolution
collisionless simulations.  \citet{kor_ken_04} pointed out that a
nuclear bar constitutes strong evidence of a pseudo-bulge, \ie, a
bulge with a disky character.  Such pseudo-bulges form through the
secular evolution of disks, via both gas and stellar dynamical
processes \citep[see the review of ][]{kor_ken_04}.  One of the main
characteristics of pseudo-bulges is that they rotate rapidly, a
property which favors them to become bar unstable.  In this work, we
demonstrate that a rapidly rotating bulge can develop a long-lived B2
in collisionless $N$-body simulations.


\section{Model Setup}
\label{sec:setup}

We focus on two examples of simulations which formed long-lasting
double-barred systems taken from a large survey of such
simulations. Our high-resolution simulations consist of live disk and
bulge components in a rigid halo potential.  We restrict ourselves to
rigid halos to afford high mass resolution in the nuclear regions, to
study the complicated co-evolution of the two bars without the
additional evolution introduced by the halo, and to compare with the
models of MS00.  We defer the study of S2B systems in live halos to a
future publication.  The rigid halos used in this study are all
logarithmic potentials $\Phi(r) = \frac{1}{2}V_{\rm h}^2~ \ln(r^2 +
r_{\rm h}^2)$.  We set $V_{\rm h} = 0.6$ in both runs and $r_{\rm h} =
15$ in run 1 and $r_{\rm h} = 10$ in run 2.
Both initial disks in our simulations have exponential surface
densities with scale-length $\Rd$, mass $\Md$ and Toomre-$Q\simeq
2$. The bulge was generated using the method of \citet{pre_tom_70} as
described in \citet{deb_sel_00}, where a distribution function is
integrated iteratively in the global potential, until convergence.  In
both cases the bulge has mass $\Mb=0.2\Md$ and we used an isotropic
King model distribution function.  The bulge truncation radius is
$0.7\Rd$ in run 1 and $1.0\Rd$ in run 2.  The bulge set up this way is
non-rotating.  We introduce bulge rotation by simply reversing the
velocities of bulge particles with negative angular momenta, which is
still a valid solution of the collisionless Boltzmann equation
\citep{lynden_62}.  The bulge in run 1 is flattened by the disk
potential to an edge-on projected ellipticity of $\epsilon_{\rm b}
\simeq 0.25$.  The ratio $V_p/{\bar\sigma} \simeq 0.8$, where $V_p$ is
the peak velocity and ${\bar\sigma}$ is the average velocity
dispersion inside the half mass radius.  In run 2, the corresponding
values are $\epsilon_{\rm b} \simeq 0.38$ and $V_p/{\bar\sigma} \simeq
0.7$.  The kinematic values relative to the oblate isotropic rotators
are $(V_p/{\bar\sigma})_* \simeq 1.3$ for run 1 and
$(V_p/{\bar\sigma})_* \simeq 0.9$ for run 2.  Thus both pseudo-bulges
are above or close to the locus of oblate isotropic rotators.  These
pseudo-bulges are less tangentially biased and more pressure supported
than rotationally-supported pseudo-bulges which would form out of gas
driven to small radii.  Our simulations therefore probably
under-estimate the tendency for pseudo-bulges to form nuclear bars.

We use $\Rd$ and $\Md$ as the units of length and mass, respectively,
and the time unit is $(\Rd^3/G\Md)^{1/2}$.  If we scale these units to
the physical values $\Md = 2.3 \times 10^{10} \Msun$ and $\Rd = 2.5$
kpc, then a unit of time is $12.3$ Myr.  We use a force resolution
(softening) of $0.01$, which scaled to the above physical units
corresponds to 25 pc.  Both models had $1.2 \times 10^6$ equal mass
particles, with $10^6$ in the disk.  These simulations were evolved
with a 3-D cylindrical polar grid code \citep{sel_val_97}.  This code
expands the potential in a Fourier series in the cylindrical polar
angle $\phi$; we truncated the expansion at $m=8$. Forces in the
radial direction are solved for by direct convolution with the Greens
function while the vertical forces are obtained by fast Fourier
transform.
We used grids measuring $N_R\times N_\phi \times N_z = 58 \times 64
\times 375$.  The vertical spacing of the grid planes was $\delta z =
0.01 R_{\rm d}$.  Time integration used a leapfrog integrator with a
fixed time-step $\delta t = 0.04$.


\section{Results}
\label{sec:results}

\begin{figure}
\centerline{
\includegraphics[angle=-90.,width=\hsize]{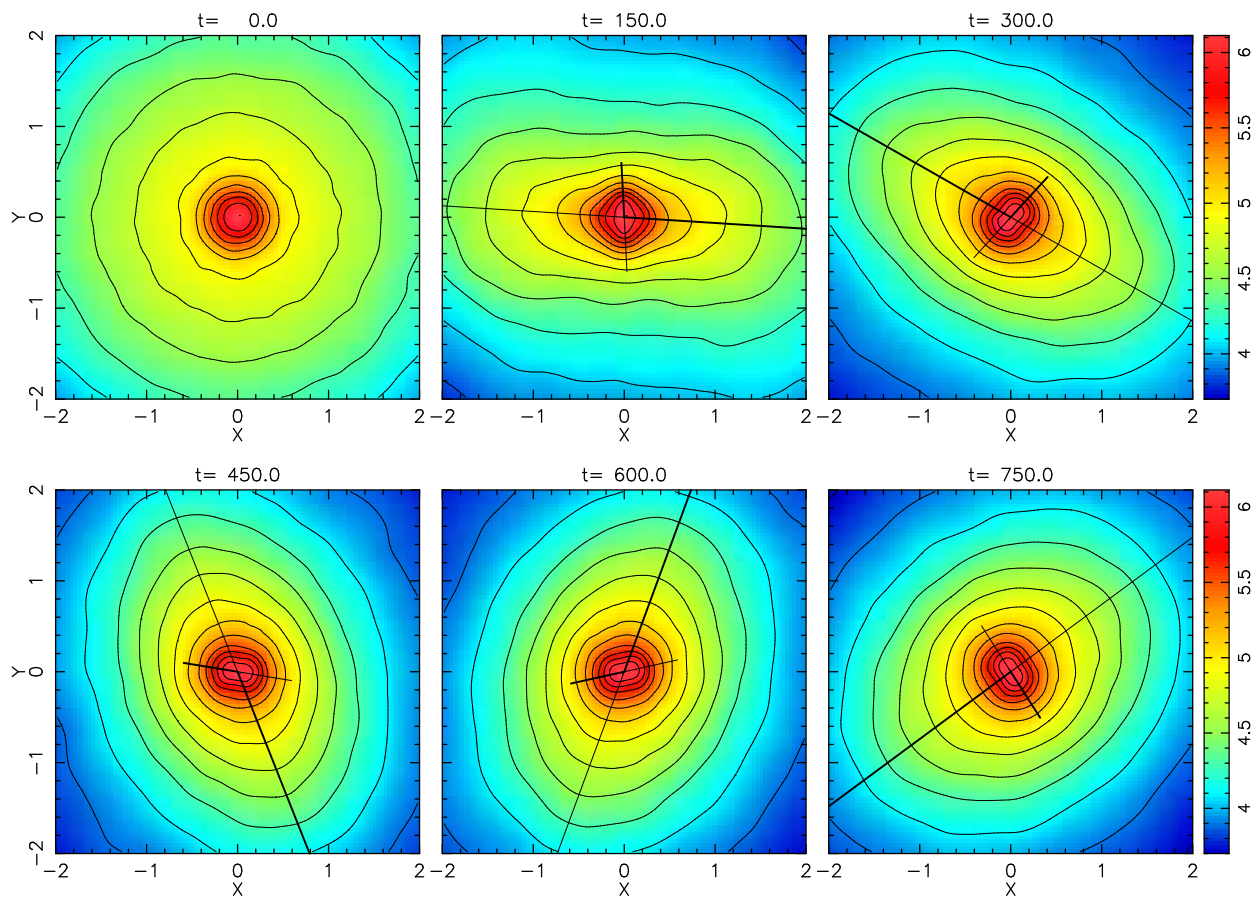}}
\caption{Images of total stellar distribution at various times, with
iso-density contours superposed. The contours are logarithmic and
separated by 0.2 dex.  The heavy short and long straight lines mark
the major axes of the B2 and B1, respectively.  The surface density is
obtained by smoothing every particle with an adaptive kernel
\citep{silver_86}.  Note that 100 time units is about 1.2 Gyr, and the
length unit is the scale-length of the initial disk.}
\label{fig:snapshots}
\end{figure}

\begin{figure}
\centerline{
\includegraphics[angle=0.,width=0.45\hsize]{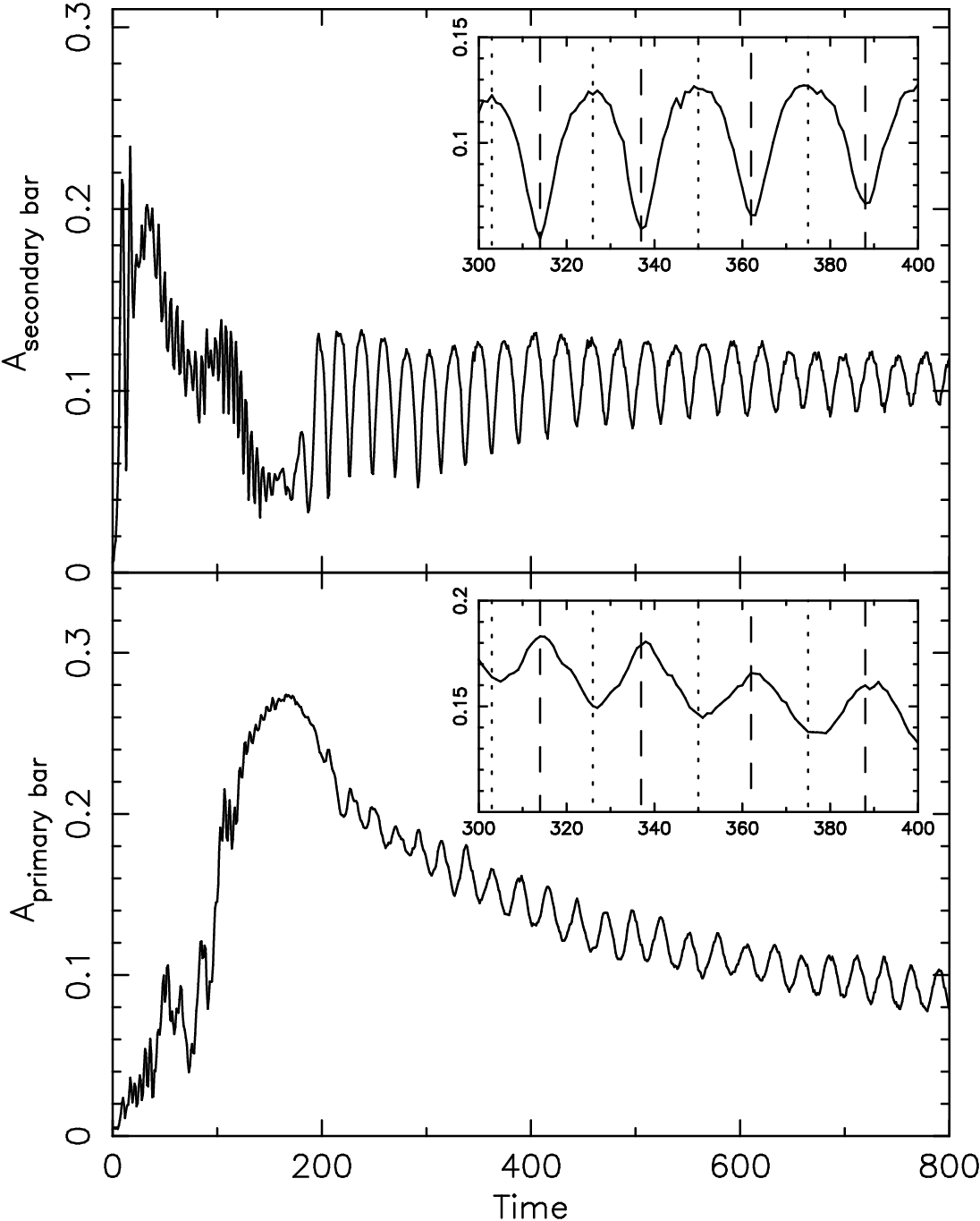}
\includegraphics[angle=0.,width=0.45\hsize]{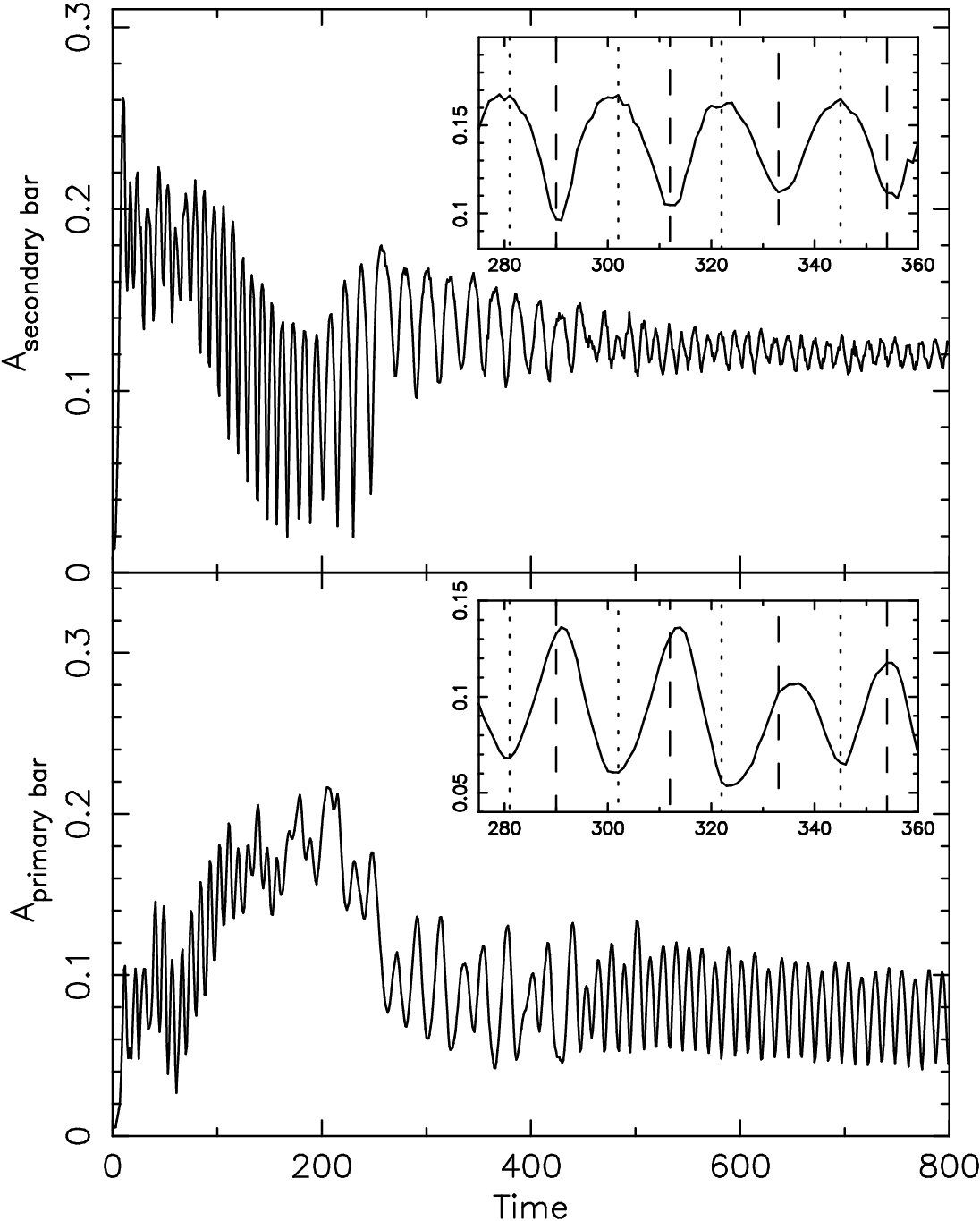}}
\caption{The time evolution of the bar amplitude of the B2 ({\it top
  panels}) and the B1 ({\it bottom panels}). In the insets, the dashed
  lines mark times when the two bars nearly align, while the dotted
  lines mark the time when they are perpendicular to each other. The
  beat period $\tme{\rm beat}=\tme{B1}\tme{B2}/2(\tme{B1}-\tme{B1})$. Run 1 is
  shown on the left, while run 2 is on the right.}
\label{fig:barampl}
\end{figure}

Figure~\ref{fig:snapshots} gives a general view of the evolution of
run 1 over 750 time units ($\sim 9.2$ Gyr in our standard scaling).  A
nuclear bar forms rapidly (before $t=10$), as the dynamical times in
the inner galaxy are much shorter than in the outer part. The pattern
speed, \omg{B2}, of this nuclear bar is large at this stage, and it
extends to nearly its corotation radius, \lag{B2}, indicating that it
forms by the usual bar instability \citep{toomre_81}.  The B1 forms at
a later stage, between $t=100$ and $200$.  The evolution of the
amplitudes of the B1 and B2 (\amp{B1} and \amp{B2} respectively),
defined as the Fourier $m=2$ amplitude over the radial ranges $ 0.5
\leq R \leq 2$ and $R \leq 0.3$, is shown in the left panel of Figure
\ref{fig:barampl}.  The B2 is strong initially, but it weakens once
the B1 forms.  At the same time $\omg{B2}$ also decreases and, since
its semi-major axis does not change substantially, it no longer
extends to \lag{B2}.
The transition to a stable S2B state is accomplished during a
seemingly chaotic period during which both bars undergo phases when
they are rather weak.  After $t=250$, however, the B2 settles into an
oscillatory steady state with \amp{B2} exhibiting regular oscillations.
The double-barred state persists to the end of the simulation, lasting
for $\sim 7$ Gyr. The B2 shows up in both the disk and bulge particles.

The B2 is stronger when the bars are perpendicular, and weaker when
they are parallel to each other (Figure \ref{fig:barampl}). Note that
this behavior is exactly opposite to the variations of gaseous rings
in \citet{hel_etal_01}. The amplitude of the primary bar instead varies
in the opposite sense with respect to the relative phase of the two
bars, although the amplitude of this oscillation is smaller.  \amp{B1}
also decreases slowly after $t=250$, possibly because orbits
supporting the B1 are gradually disrupted by the relatively strong
inner bar.  As a consequence, the oscillations in \amp{B2} decrease as
the B1 weakens.

The right panel of Figure \ref{fig:barampl} shows the evolution of
\amp{B1} and \amp{B2} in run 2.  The main difference between run 2 and
run 1 is that the initial bulge is larger in run 2, allowing the B2 to
dominate the global dynamics.  As a result, the B1 oscillates more
strongly than the B2.  This probably represents an extreme case of the
dynamical influence of a B2 on a B1. 

\begin{figure}
\centerline{
\includegraphics[angle=-90., width=\hsize]{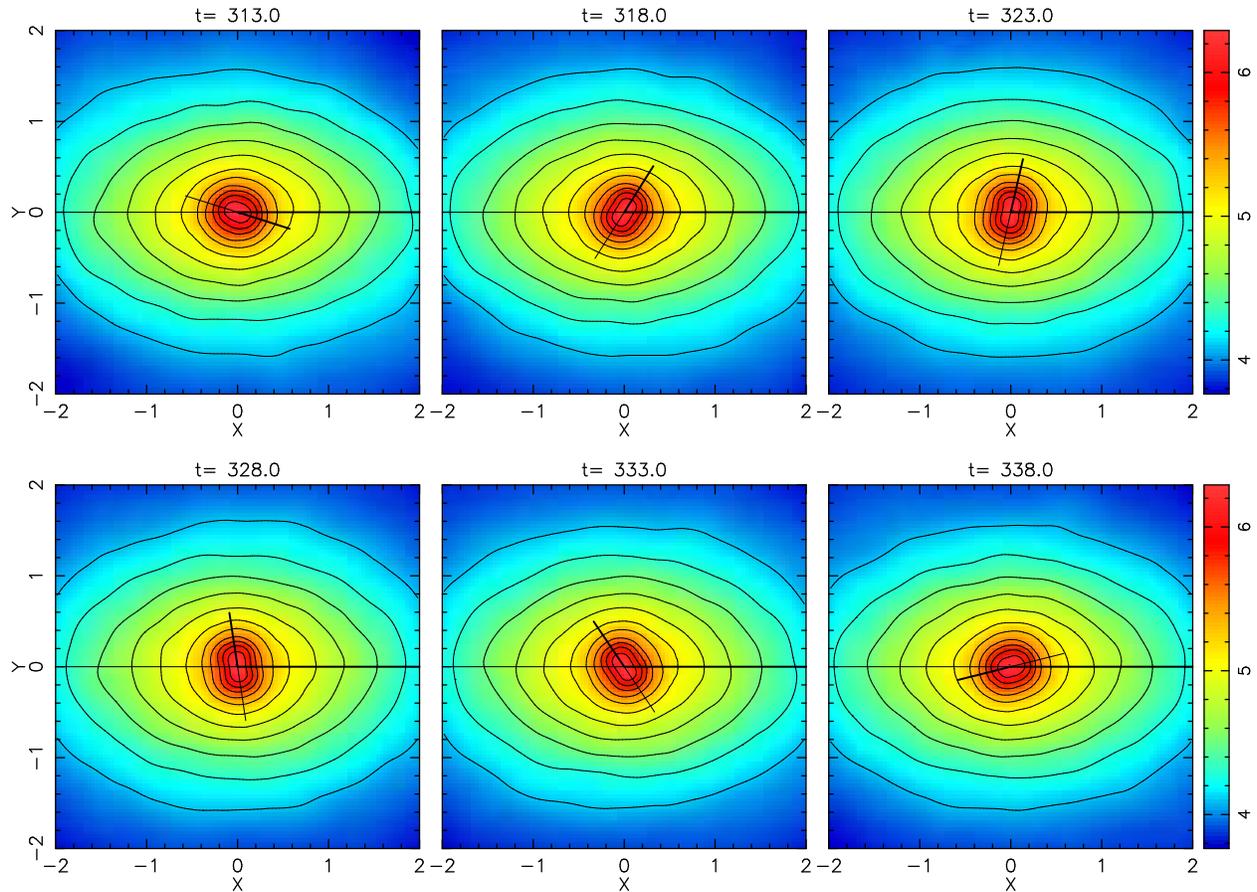}}
\caption{The non-uniform relative rotation of the B2 for roughly half
  of a period, in the corotating frame of the B1 which remains
  horizontal.  The panels are equally-spaced in time. The straight
  line marks the major axis of the B2.  The B2 rotates faster when the
  two bars align than when the two bars are perpendicular.}
\label{fig:crframe}
\end{figure}

\begin{figure}
\centerline{
\includegraphics[angle=-90.,width=\hsize]{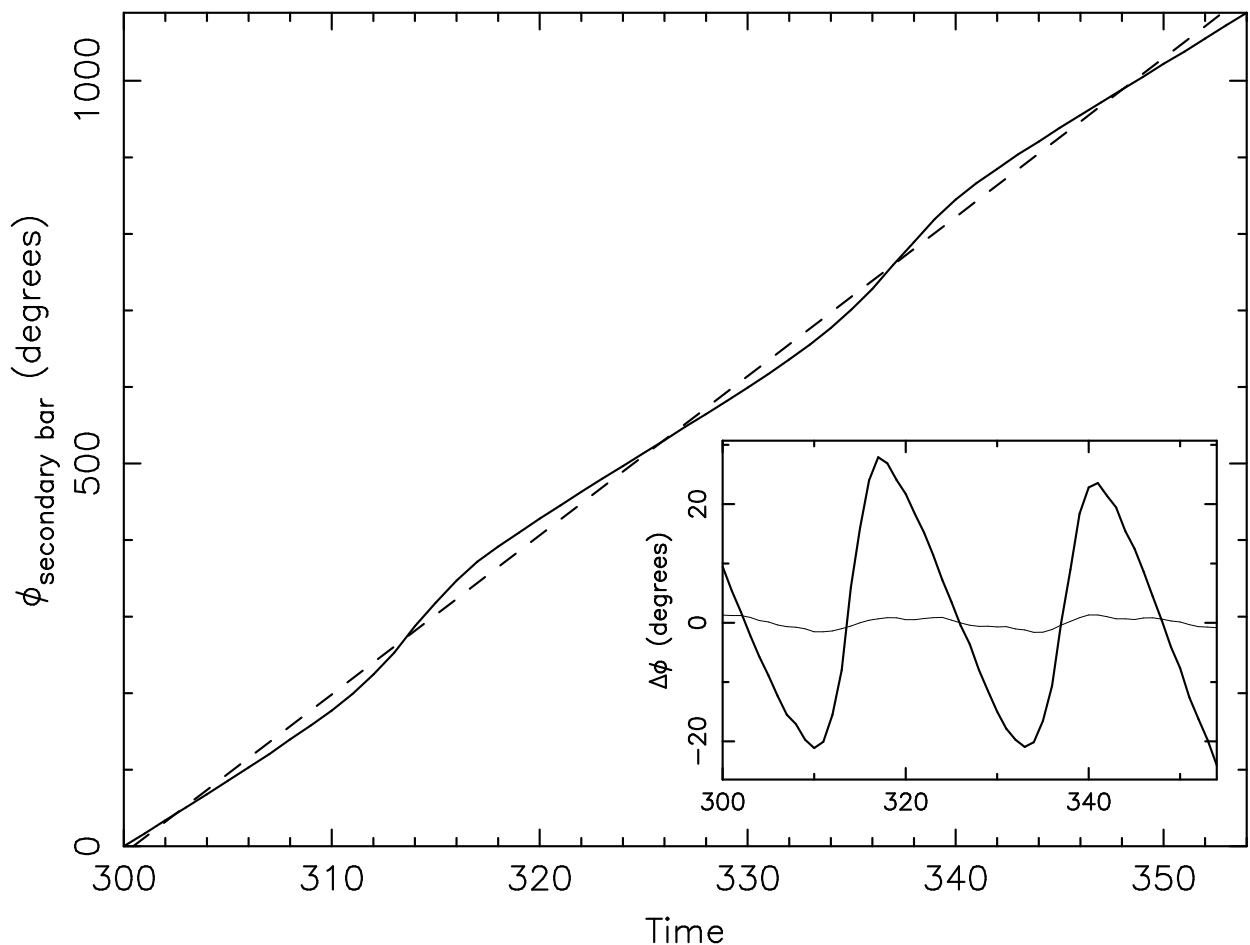}}
\caption{The time evolution of the phase of the B2, measured relative
to $t=300$. The dashed straight line is the least-square fit which
gives the slope $\left<\omg{B2}\right>$. The inset figure shows the
phase difference, $\Delta\phi$ between the phases of the bars and
$\left<\omg{}\right>t$, where $\left<\omg{}\right>$ is the pattern
speed averaged over one relative rotation of the two bars; the thick
line is for the B2 while the thin line is for the B1.}
\label{fig:barphase}
\end{figure}

The long-lived B2 rotates faster than the B1: between $t=300$ and
$t=400$ the average rotation period of the B2 in run 1 is about
$\tme{B2}\simeq 17.6$, and for the B1 $\tme{B1} \simeq 27.8$.  The
pattern speed of the B2, \omg{B2}, also varies with the relative phase
of the two bars: it is larger when the two bars align, and smaller
when they are orthogonal. We plot in Figure \ref{fig:crframe} the
system in the corotating frame of the B1.  The variations of both
\amp{B2} and \omg{B2} are readily visible. The variation of \omg{B2}\ can
be $>20\%$ but is much less significant for \omg{B1} (Figure
\ref{fig:barphase}).  Defining $\left< \omg{B2}\right>$ ($\left<
\omg{B1}\right>$) as the average pattern speed of the B2 (B1) over one
relative rotation, we plot in the inset of Figure \ref{fig:barphase}
the phase difference between $\left< \omg{B2}\right>t$ ($\left<
\omg{B1}\right>t$) and the phase of the B2 (B1).  The B1 is seen to
rotate with a rather constant \omg{B1}\ but the B2 experiences a large
variation in \omg{B2} over one relative rotation.

\begin{figure}
\centerline{
\includegraphics[angle=-90.,width=0.5\hsize]{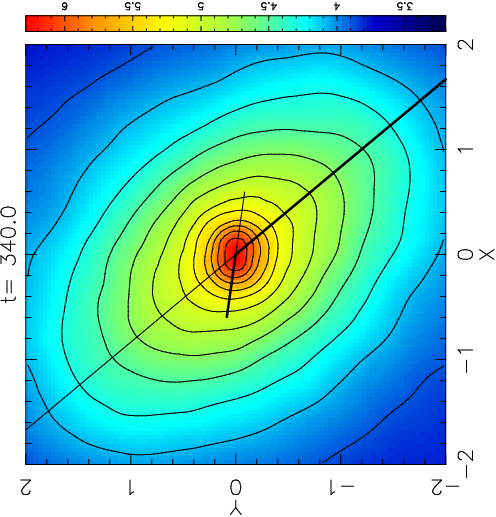}
\includegraphics[angle=-90.,width=0.5\hsize]{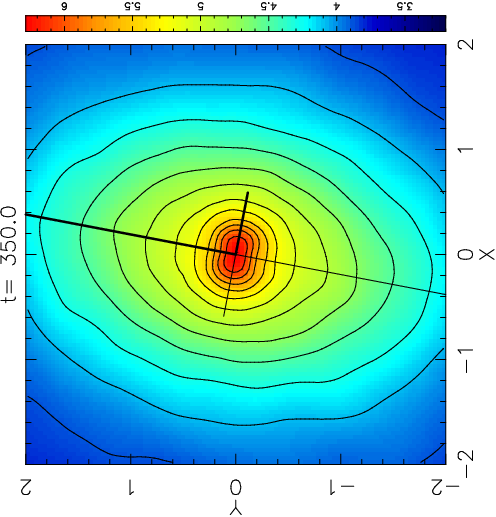}}
\centerline{
\includegraphics[angle=0.,width=0.5\hsize]{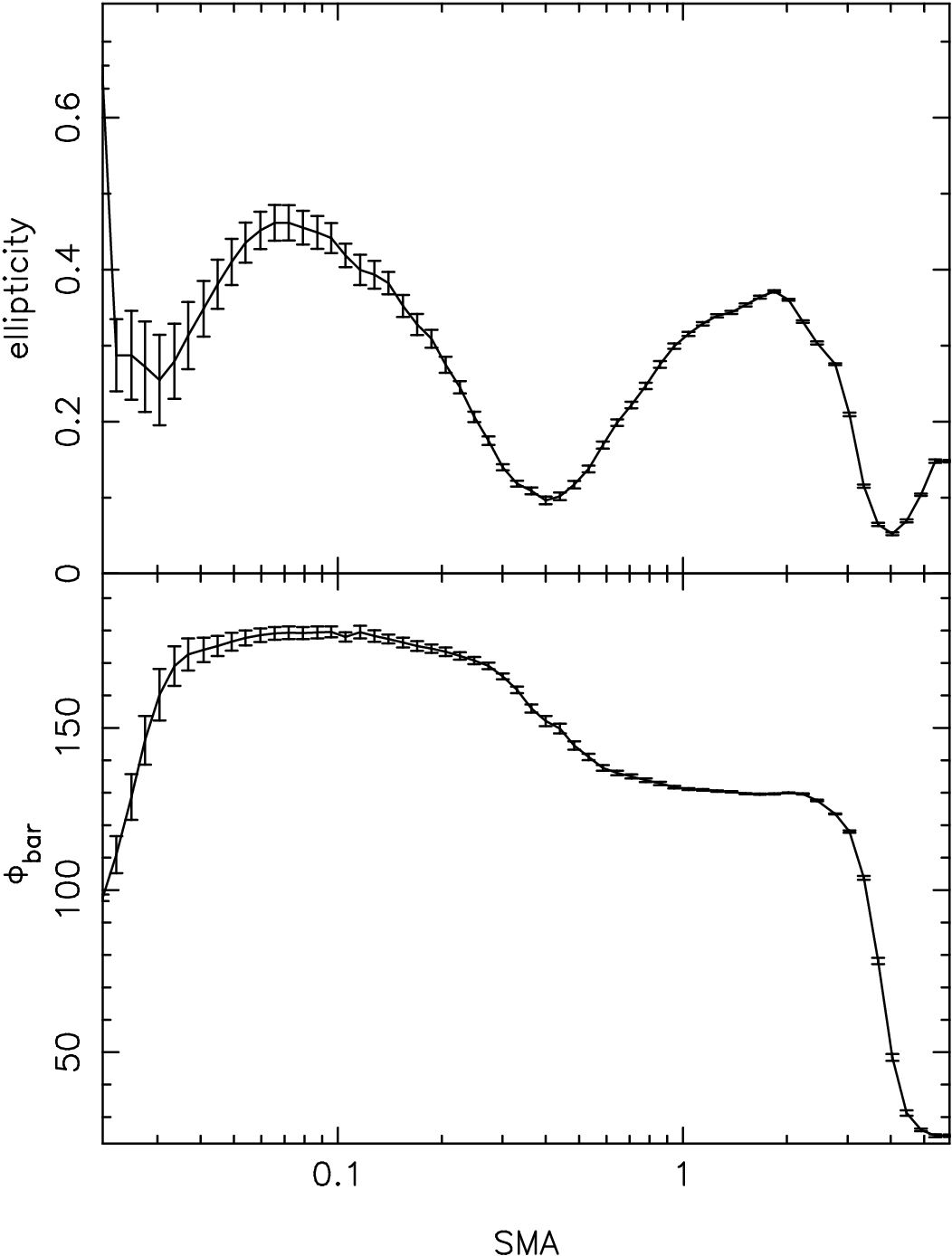}
\includegraphics[angle=0.,width=0.5\hsize]{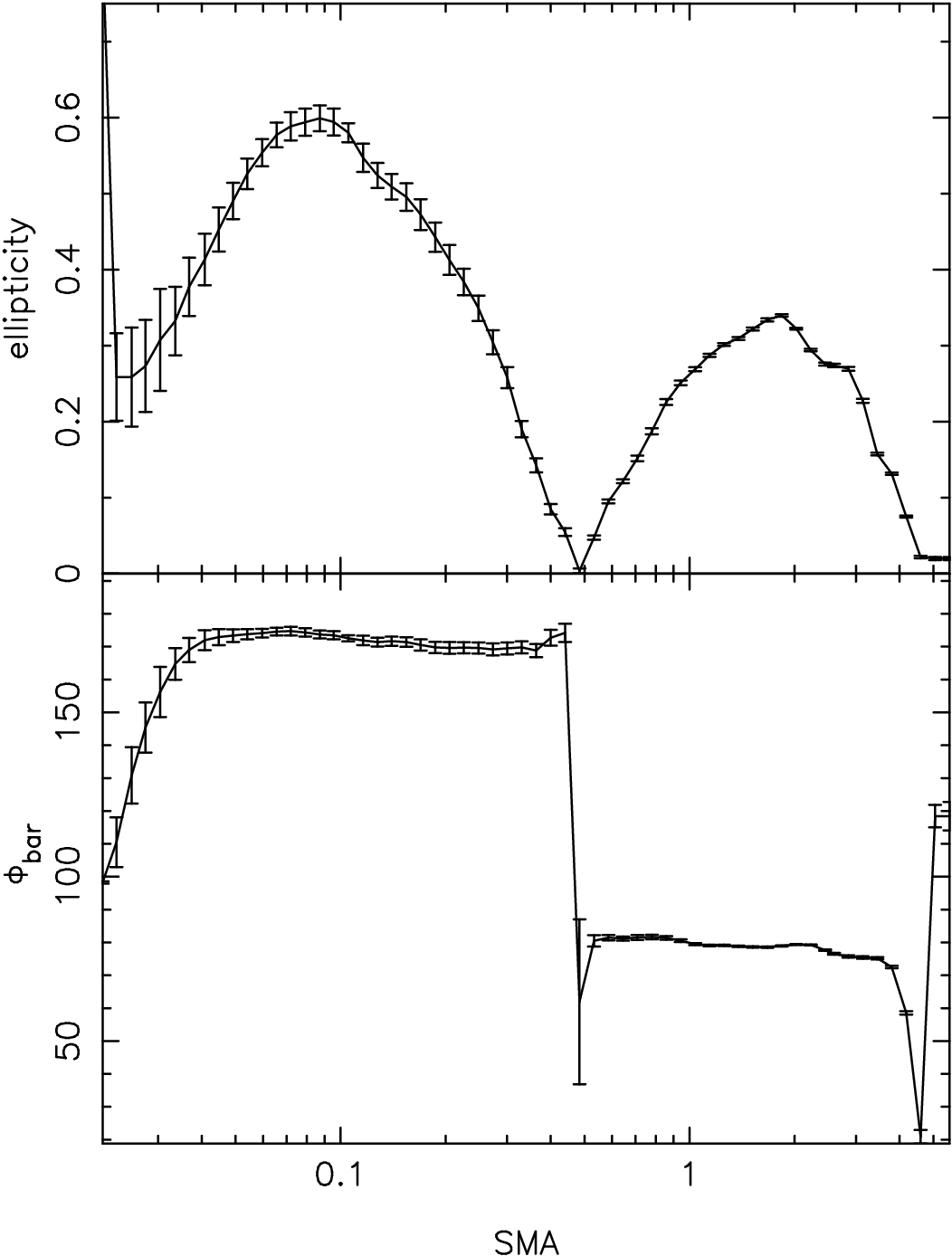}}
\caption{Results of ellipse fits using {\sc iraf}.  Top panels:
snapshots of run 1 at $t=340$ (left) and $t=350$ (right).  The two
bars are at $\sim 45\degrees$ at $t=340$ and perpendicular at $t=350$.
Middle panels: Ellipticity as a function of semi-major axis (SMA) of
fitted ellipses.  Bottom panels: Position angle as a function of SMA
of fitted ellipses.}
\label{fig:ellipsefit}
\end{figure}

Figure \ref{fig:ellipsefit} presents ellipse fits using {\sc iraf} for
times when the B2 and B1 are perpendicular and at $\sim 45\degrees$ to
each other.  In both cases the phase of the B2 is constant to within
$10\degrees$ and there is little sense of spirality in it.  This is
distinctly a nuclear bar rather than a nuclear spiral.

We measured the sizes of the two bars, for two different relative
orientations at $t=340$ and at $t=350$, as the larger radius where the
bar phase deviates by more than $10\degrees$ from a constant value.
We find a semi-major axis ratio $\simeq 0.10$ ($\simeq 0.12$) at
$t=340$ ($t=350$), in good agreement with the typical size ratio of local
S2B systems \citep{erw_spa_02}.


\section{Discussion and Conclusions}
\label{sec:discussion}

Our self-consistent simulations of S2B systems can be compared to the
models of MS00.  The simulations all exhibited oscillating pattern
speeds and amplitudes for one or both bars.  Similarly MS00 found that
the $x_2$ loops supporting the B2 change axis ratios and lead or trail
the rigid figure of the B2, as the bars rotate.  The loop orbits of
MS00 were more elongated in the B2 region when the two bars were
orthogonal than when they were parallel.  The pulsating character of
the self-consistent B2 in the simulations provides strong evidence
that $x_2$ loops are the backbone of the double bars in these
simulations.  This behavior is also in good agreement with the earlier
prediction by \citet{lou_ger_88} that independent rigid rotation of
two bars is not possible.
The $x_2$ loop orbits of MS00 also suggested that \omg{B2} would be
largest when the two bars are parallel, which is also borne out by the
simulations.  Furthermore, MS00 were unable to find supporting $x_2$
orbits when the B2 extended to about \lag{B2}, while we found that
\omg{B2} had to decrease once the B1 formed and the B2 did not extend
to \lag{B2}, again in good agreement with MS00. 
Our simulations also suggest that observationally there should be a
slight excess of close-to-perpendicular double bars, as the secondary
bar tends to rotate slower when two bars are perpendicular.

The main objective of this work is to create S2B systems and show how
the two bars form spontaneously, interact and evolve.  Our simulations
all form B2s before they form B1s.  However this is not a prediction
of our model and it occurs only because, for simplicity, we introduced
our rotating pseudo-bulge from $t=0$.  It is more likely that a
pseudo-bulge would form after gas is driven to the center by a
pre-existing B1.  Our pseudo-bulges all had rotation; B2s did not form
in simulations without pseudo-bulge rotation \citep[these results will
be presented elsewhere, but see for example][]{debatt_03}.  In
contrast, \citet{rau_sal_99} produced B2s even though their bulges
were analytic.

We are able to form well-resolved, long-lived B2s in purely
collisionless $N$-body simulations.  The nuclear bars are distinctly
barred, not spiral, and reach to the center.  These simulations
demonstrate that B2s do not need to be gaseous.  We confirm that
pseudo-bulge rotation may be an important ingredient for the formation
of double-barred galaxies \citep{kor_ken_04}. The required degree of
rotation is modest and not greater than that associated with
pseudo-bulges \citep{kormen_93}.  The B2s in these simulations rotate
faster than the B1s.  The implications of the pulsating nature of B1s
on central gas inflow are unclear at present.
This new method for forming S2B models reliably and repeatedly should
prove a boon to exploring their dynamics and evolution, their
observational properties, their effect on gas, \etc\ We will report on
these issues elsewhere.

\acknowledgements 
V.P.D. is supported by a Brooks Prize Fellowship at the
University of Washington and receives partial support from NSF ITR
grant PHY-0205413.  V.P.D. thanks the University of Texas at Austin
for hospitality during part of this project.


\clearpage

\clearpage

\clearpage

\clearpage

\clearpage

\begin{thebibliography}{}

\bibitem[\protect\citeauthoryear{{Begelman}, {Volonteri}, \& {Rees}}{{Begelman}
  et~al.}{2006}]{beg_etal_06}
{Begelman}, M.~C., {Volonteri}, M.,  \& {Rees}, M.~J. 2006, astro-ph/0602363

\bibitem[\protect\citeauthoryear{{Buta} \& {Crocker}}{{Buta} \&
  {Crocker}}{1993}]{but_cro_93}
{Buta}, R.,  \& {Crocker}, D.~A. 1993, \aj, 105, 1344

\bibitem[\protect\citeauthoryear{{Corsini}, {Debattista}, \&
  {Aguerri}}{{Corsini} et~al.}{2003}]{cor_deb_agu_03}
{Corsini}, E.~M., {Debattista}, V.~P.,  \& {Aguerri}, J.~A.~L. 2003, \apjl,
  599, L29

\bibitem[\protect\citeauthoryear{{Davies} \& {Hunter}}{{Davies} \&
  {Hunter}}{1997}]{dav_hun_97}
{Davies}, C.~L.,  \& {Hunter}, J.~H. 1997, \apj, 484, 79

\bibitem[\protect\citeauthoryear{{de Vaucouleurs}}{{de
  Vaucouleurs}}{1975}]{devauc_75}
{de Vaucouleurs}, G. 1975, \apjs, 29, 193

\bibitem[\protect\citeauthoryear{{Debattista}}{{Debattista}} {2003}]{debatt_03}
{Debattista}, V.~P. 2003, \mnras, 342, 1194

\bibitem[\protect\citeauthoryear{{Debattista} \& {Sellwood}}{{Debattista} \&
  {Sellwood}}{2000}]{deb_sel_00}
{Debattista}, V.~P.,  \& {Sellwood}, J.~A. 2000, \apj, 543, 704

\bibitem[\protect\citeauthoryear{{Emsellem} et~al.}{{Emsellem}
  et~al.}{2001}]{ems_etal_01}
{Emsellem}, E., {Greusard}, D., {Combes}, F., {Friedli}, D., {Leon}, S.,
  {P{\'e}contal}, E.,  \& {Wozniak}, H. 2001, \aap, 368, 52

\bibitem[\protect\citeauthoryear{{Englmaier} \& {Shlosman}}{{Englmaier} \&
  {Shlosman}}{2004}]{eng_shl_04}
{Englmaier}, P.,  \& {Shlosman}, I. 2004, \apjl, 617, L115

\bibitem[\protect\citeauthoryear{{Erwin}}{{Erwin}}{2004}]{erwin_04}
{Erwin}, P. 2004, \aap, 415, 941

\bibitem[\protect\citeauthoryear{{Erwin} \& {Sparke}}{{Erwin} \&
  {Sparke}}{2002}]{erw_spa_02}
{Erwin}, P.,  \& {Sparke}, L.~S. 2002, \aj, 124, 65

\bibitem[\protect\citeauthoryear{{Friedli}}{{Friedli}}{1996}]{friedl_96}
{Friedli}, D. 1996, \aap, 312, 761

\bibitem[\protect\citeauthoryear{{Friedli} \& {Martinet}}{{Friedli} \&
  {Martinet}}{1993}]{fri_mar_93}
{Friedli}, D.,  \& {Martinet}, L. 1993, \aap, 277, 27

\bibitem[\protect\citeauthoryear{{Heller}, {Shlosman}, \& {Englmaier}}{{Heller}
  et~al.}{2001}]{hel_etal_01}
{Heller}, C., {Shlosman}, I.,  \& {Englmaier}, P. 2001, \apj, 553, 661

\bibitem[\protect\citeauthoryear{{Kormendy}}{{Kormendy}}{1993}]{kormen_93}
{Kormendy}, J. 1993, in IAU Symp. 153: Galactic Bulges, ed. H.~{Dejonghe} \&
  H.~J. {Habing}, 209

\bibitem[\protect\citeauthoryear{{Kormendy} \& {Kennicutt}}{{Kormendy} \&
  {Kennicutt}}{2004}]{kor_ken_04}
{Kormendy}, J.,  \& {Kennicutt}, R.~C. 2004, \araa, 42, 603

\bibitem[\protect\citeauthoryear{{Kuijken}, {Fisher}, \&
  {Merrifield}}{{Kuijken} et~al.}{1996}]{kui_etal_96}
{Kuijken}, K., {Fisher}, D.,  \& {Merrifield}, M.~R. 1996, \mnras, 283, 543

\bibitem[\protect\citeauthoryear{{Laine} et~al.}{{Laine}
  et~al.}{2002}]{lai_etal_02}
{Laine}, S., {Shlosman}, I., {Knapen}, J.~H.,  \& {Peletier}, R.~F. 2002, \apj,
  567, 97

\bibitem[\protect\citeauthoryear{{Louis} \& {Gerhard}}{{Louis} \&
  {Gerhard}}{1988}]{lou_ger_88}
{Louis}, P.~D.,  \& {Gerhard}, O.~E. 1988, \mnras, 233, 337

\bibitem[\protect\citeauthoryear{{Lynden-Bell}}{{Lynden-Bell}}{1962}]{lynden_62}
{Lynden-Bell}, D. 1962, \mnras, 123, 447

\bibitem[\protect\citeauthoryear{{Maciejewski} \& {Sparke}}{{Maciejewski} \&
  {Sparke}}{1997}]{mac_spa_97}
{Maciejewski}, W.,  \& {Sparke}, L.~S. 1997, \apjl, 484, L117

\bibitem[\protect\citeauthoryear{{Maciejewski} \& {Sparke}}{{Maciejewski} \&
  {Sparke}}{2000}]{mac_spa_00}
{Maciejewski}, W.,  \& {Sparke}, L.~S. 2000, \mnras, 313, 745

\bibitem[\protect\citeauthoryear{{Maciejewski} et~al.}{{Maciejewski}
  et~al.}{2002}]{mac_etal_02}
{Maciejewski}, W., {Teuben}, P.~J., {Sparke}, L.~S.,  \& {Stone}, J.~M. 2002,
  \mnras, 329, 502

\bibitem[\protect\citeauthoryear{{Moiseev}, {Vald{\'e}s}, \&
  {Chavushyan}}{{Moiseev} et~al.}{2004}]{moi_etal_04}
{Moiseev}, A.~V., {Vald{\'e}s}, J.~R.,  \& {Chavushyan}, V.~H. 2004, \aap, 421,
  433

\bibitem[\protect\citeauthoryear{{Mulchaey}, {Regan}, \& {Kundu}}{{Mulchaey}
  et~al.}{1997}]{mul_etal_97}
{Mulchaey}, J.~S., {Regan}, M.~W.,  \& {Kundu}, A. 1997, \apjs, 110, 299

\bibitem[\protect\citeauthoryear{{Petitpas} \& {Wilson}}{{Petitpas} \&
  {Wilson}}{2002}]{pet_wil_02}
{Petitpas}, G.~R.,  \& {Wilson}, C.~D. 2002, \apj, 575, 814

\bibitem[\protect\citeauthoryear{{Prendergast} \& {Tomer}}{{Prendergast} \&
  {Tomer}}{1970}]{pre_tom_70}
{Prendergast}, K.~H.,  \& {Tomer}, E. 1970, \aj, 75, 674

\bibitem[\protect\citeauthoryear{{Rautiainen}, \& {Salo}}{{Rautiainen}
\& {Salo}}{1999}]{rau_sal_99} {Rautiainen}, P., \& {Salo}, H.  1999,
\aap, 348, 737

\bibitem[\protect\citeauthoryear{{Rautiainen}, {Salo}, \&
  {Laurikainen}}{{Rautiainen} et~al.}{2002}]{rau_etal_02}
{Rautiainen}, P., {Salo}, H.,  \& {Laurikainen}, E. 2002, \mnras, 337, 1233

\bibitem[\protect\citeauthoryear{{Schinnerer} et~al.}{{Schinnerer}
  et~al.}{2002}]{sch_etal_2002}
{Schinnerer}, E., {Maciejewski}, W., {Scoville}, N.,  \& {Moustakas}, L.~A.
  2002, \apj, 575, 826

\bibitem[\protect\citeauthoryear{{Sellwood} \& {Merritt}}{{Sellwood} \&
  {Merritt}}{1994}]{sel_mer_94}
{Sellwood}, J.~A.,  \& {Merritt}, D. 1994, \apj, 425, 530

\bibitem[\protect\citeauthoryear{{Sellwood} \& {Valluri}}{{Sellwood} \&
  {Valluri}}{1997}]{sel_val_97}
{Sellwood}, J.~A.,  \& {Valluri}, M. 1997, \mnras, 287, 124

\bibitem[\protect\citeauthoryear{{Shlosman}, {Frank}, \& {Begelman}}{{Shlosman}
  et~al.}{1989}]{shl_etal_89}
{Shlosman}, I., {Frank}, J.,  \& {Begelman}, M.~C. 1989, \nat, 338, 45

\bibitem[\protect\citeauthoryear{{Shlosman} \& {Heller}}{{Shlosman} \&
  {Heller}}{2002}]{shl_hel_02}
{Shlosman}, I.,  \& {Heller}, C.~H. 2002, \apj, 565, 921

\bibitem[\protect\citeauthoryear{{Silverman}}{{Silverman}}{1986}]{silver_86}
{Silverman}, B.~W. 1986, {Density estimation for statistics and data analysis}
  (Monographs on Statistics and Applied Probability, London: Chapman and Hall)

\bibitem[\protect\citeauthoryear{{Springel}, {Di Matteo}, \&
  {Hernquist}}{{Springel} et~al.}{2005}]{spr_etal_05b}
{Springel}, V., {Di Matteo}, T.,  \& {Hernquist}, L. 2005, \mnras, 361, 776

\bibitem[\protect\citeauthoryear{{Toomre}}{{Toomre}}{1981}]{toomre_81}
{Toomre}, A. 1981, in Structure and Evolution of Normal Galaxies, ed. S.~M.,
  Fall \& D. Lynden-Bell (Cambridge: Cambridge University Press), 111

\bibitem[\protect\citeauthoryear{{Tremaine} \& {Weinberg}}{{Tremaine} \&
  {Weinberg}}{1984}]{tre_wei_84}
{Tremaine}, S.,  \& {Weinberg}, M.~D. 1984, \apjl, 282, L5

\end{thebibliography}
\end{document}